\documentclass[10pt]{article}
\usepackage[margin=0.65in]{geometry}
\usepackage[utf8]{inputenc}
\usepackage{aaai}
\usepackage{times}
\usepackage{helvet}
\usepackage{verbatim}
\usepackage{courier}
\usepackage[pdftex]{graphicx}
\usepackage[hyphens]{url}
\usepackage{subcaption,caption}
\usepackage{float}
\nocopyright

\begin{document}

\title{Partisanship, Propaganda and Post-Truth Politics: Quantifying Impact in Online Debate}

\date{}

\author{
  Genevieve Gorrell, 
  Mehmet E. Bakir, 
  Ian Roberts, 
  \\
  \Large{\textbf{Mark A. Greenwood, 
  Benedetta Iavarone and
  Kalina Bontcheva}}
  \\
  \\
  \Large{\textbf{University of Sheffield, UK}}
}

\maketitle

\begin{abstract}
  The recent past has highlighted the influential role of social networks and online media in shaping public debate on current affairs and political issues. This paper is focused on studying the role of politically-motivated actors and their strategies for influencing and manipulating public opinion online: partisan media, state-backed propaganda, and post-truth politics. In particular, we present quantitative research on the presence and impact of these three ``Ps'' in online Twitter debates in two contexts: (i) the run up to the UK EU membership referendum (``Brexit''); and (ii) the information operations of Russia-backed online troll accounts. We first compare the impact of highly partisan versus mainstream media during the Brexit referendum, specifically comparing tweets by half a million ``leave'' and ``remain'' supporters. Next, online propaganda strategies are examined, specifically left- and right-wing troll accounts.  Lastly, we study the impact of misleading claims made by the political leaders of the leave and remain campaigns.  This is then compared to the impact of the Russia-backed partisan media and propaganda accounts during the referendum. In particular, just two of the many misleading claims made by politicians during the referendum were found to be cited in 4.6 times more tweets than the 7,103 tweets related to Russia Today and Sputnik and in 10.2 times more tweets than the 3,200 Brexit-related tweets by the Russian troll accounts.
\end{abstract}

\section{Introduction}

``Post-truth politics''~\cite{higgins2016post} and ``weaponized
relativism''\footnote{\small{\url{https://www.theguardian.com/commentisfree/2015/mar/02/guardian-view-russian-propaganda-truth-out-there}}}
describe strategies by which misleading information can be used to
shape debates, redirect attention  and sow confusion in order to
influence political outcomes. In recent times, concern has been raised
about  politicians, foreign states, and hyper-partisan media exploiting social media to try and reach out and influence voters and citizens on an unprecedented scale. Where once social media were heralded as the beginning of a new age of interactive democracy, the question in the minds of researchers and many others is now ``can democracy survive the
internet''~\cite{persily20172016}. A working theory might postulate that the low bar to publishing created by Web 2.0 has resulted in a number of effects that we explore here under three headings:

\begin{itemize}

\item{\textbf{Partisan media}: today’s highly competitive online media
   landscape has resulted in poorer quality journalism and worsening
   opinion diversity, with misinformation, bias and factual
   inaccuracies routinely creeping in. Many outlets also resort to
   highly partisan reporting of key political events, which can have
   acrimonious and divisive effects.}

\item{Online \textbf{propaganda}: State-backed
   (e.g. Russia Today), ideology-driven (e.g.  misogynistic or
   Islamophobic), or for-profit clickbait websites and social media
   accounts are engaged in spreading manipulative content and disinformation often with the intent to deepen social division and/or influence key political outcomes.}

\item{\textbf{Post-truth politics}, where politicians, parties and
   governments frame key political issues in propaganda instead of
   facts. Misleading claims are repeated, even when proven untrue by
   journalists or independent fact checkers. This has a highly corrosive
   effect on public trust and informed participation in democratic
processes.}

\end{itemize}

While researchers have started studying these
recently~\cite{skjesethall,ferrara2017disinformation}, the majority of
work has focused primarily on misinformation and fake news during
elections~\cite{vosoughi2018spread,kaminskajunk} and the role of bots in
spreading it \cite{shao2018spread,howard2016bots}. This paper presents large-scale, quantitative research on the presence and impact of these three ``Ps'' in online Twitter debates in two contexts: (i) the run up to the UK EU membership referendum (``Brexit''); and (ii) the information operations of Russia-backed online troll accounts.  The aggregate data on which this work is based will be made available online upon publication.

We first compare the impact of highly partisan versus mainstream media
during the Brexit referendum, specifically comparing tweets by half a
million ``leave'' and ``remain'' supporters. Next, online propaganda strategies
are examined, specifically differentiating left- and right-wing troll
accounts.  Lastly, we study the impact of misleading claims made by the
political leaders of the leave and remain campaigns.  This is then
compared to the impact of the Russia-backed partisan media and
propaganda accounts during the referendum. In particular, just two of
the many misleading claims made by politicians during the referendum
were found to be cited in 4.6 times more tweets than the 7,103 tweets
related to Russia Today and Sputnik and in 10.2 times more tweets than
the 3,200 Brexit-related tweets by the Russian troll accounts.

Furthermore, late in 2018 Twitter released a set of nine million tweets
from accounts they have identified as belonging to the Russian Internet
Research Agency (IRA). The IRA dataset covers a time period spanning
from the beginning of the Ukraine conflict in 2014 through the Brexit
referendum and US presidential election until well into President Donald
Trump's term of office. These data provide rich possibilities for
investigating propaganda. We present here the first exhaustive analysis
of this new dataset, with a focus on what we can learn about how
propaganda succeeds and fails under the conditions created by modern
social media. We also accurately classify accounts into different
activity types (left trolls, right trolls, etc.), enabling a deeper
understanding of how different strategies pay off in terms of impact.

\subsection{Related Work}

The work presented here is set against a backdrop of increasing
awareness of the ways in which the internet and social media are
changing society. Social media have been widely observed to provide a
platform for fringe views. Faris \textit{et
al}~\shortcite{faris2017partisanship} showed that social media seem to
amplify more extreme views, with materials linked on Twitter being more
outr\'e than the open web, and on Facebook even more so, a finding
echoed by Silverman~\shortcite{silverman2015lies}. Barber\'a and
Rivero~\shortcite{barbera2015understanding} and Preotiuc-Pietro \textit{et
al}~\shortcite{preoctiuc2017beyond} both show that Twitter users with more
ideologically extreme positions post more content than those with
moderate views.

Researchers also report consistent asymmetries in the way these changed
conditions play out. Allcott and Gentzkow~\shortcite{allcott2017social},
during the run-up to the 2016 US presidential election, found 115
pro-Trump fake news stories, which were shared a total of 30 million
times. They found 41 pro-Clinton fake news stories, which were shared a
total of 7.6 million times. This disparity is again echoed in
Silverman's~\shortcite{silverman2015lies} work. Hare and
Poole~\shortcite{hare2014polarization} find that the increased separation
between American left and right wing partisans in recent years is
accounted for by a right wing shift to the right; left wing voters have
not changed their position.

There is little evidence of a difference in the way information
consumers of different political valences respond to materials that
might account for this
asymmetry~\cite{faris2017partisanship,allcott2017social}. Instead, Faris
\textit{et al} suggest that in the case of the 2016 presidential
election, it was the cooperative behaviour of pro-Trump media themselves
that led to an advantage, in a phenomenon they dub ``network
propaganda''. This raises questions about the reach of such a network or
the conditions under which it might arise elsewhere, and its
relationship to political views if any. The idea of an ``alternate
reality'' created by network propaganda has implications for social
polarization given Lewandowsky \textit{et
al}'s~\shortcite{lewandowsky2017beyond} observation that where partisans are
isolated in echo chambers extremism is rewarded, as a message may reach
sympathizers without the cost attached in alienating centric or opposing
voters.

A body of work~\cite{lansdall2016change,mangold2016should} has begun to
explore Brexit opinion and sentiment as expressed on Twitter. Matsuo and
Benoit~\footnote{\small{\url{http://blogs.lse.ac.uk/brexit/2017/03/16/more-positive-assertive-and-forward-looking-how-leave-won-twitter/}}}
focus on differences in the dialogue between leave and remain camps.
Moore and Ramsay~\shortcite{moore2017uk}'s mostly manual research is focused
on analysing the newspaper media during the referendum and highlights
differences in the tone of the different campaigns.  Our work builds on
theirs by exploring how the behaviour they discuss relates to a medium's
partisan appeal, as well as focusing on social media, rather than
newspapers.

Howard and Kollanyi~\shortcite{howard2016bots} share our interest in
propaganda. Our novel contributon is in exploiting large-scale, reliable
voter classification in order to explore partisan dynamics and
polarisation. Their group have also specifically investigated Russian
bot involvement in Brexit~\cite{narayananrussian}, but on a
significantly smaller scale. Likewise, Bastos and
Mercea~\shortcite{bastos2017brexit} study the impact of bot activity during
Brexit, and present some observations about the nature of the content
they spread. They find that such materials are likely to be
user-generated, tabloid-style emotionally orientated materials. The role
of Twitter misinformation and bot activity in the context of the 2016 US
presidential election has attracted much research attention, as
previously discussed. This has primarily focused on the amount of
traffic generated by bots or trolls, without providing evidence of
impact. In this paper, instead, we focus on quantifying bot impact and
investigating the strategies for achieving it.

The release of the IRA dataset is so recent as to preclude much in the
way of in-depth investigation so far.The largest prior study available
by Linvill and Warren~\shortcite{linvill2018troll} still had access to only 3
million tweets, which is very significantly less than the 9 million just
released by Twitter. This new large corpus constitutes an unprecedented
opportunity, since troll accounts are rapidly suspended by the platform,
creating a moving target for research.

\subsection{Term Definitions}

\begin{figure}
  \centering
\includegraphics[width=0.40\textwidth]{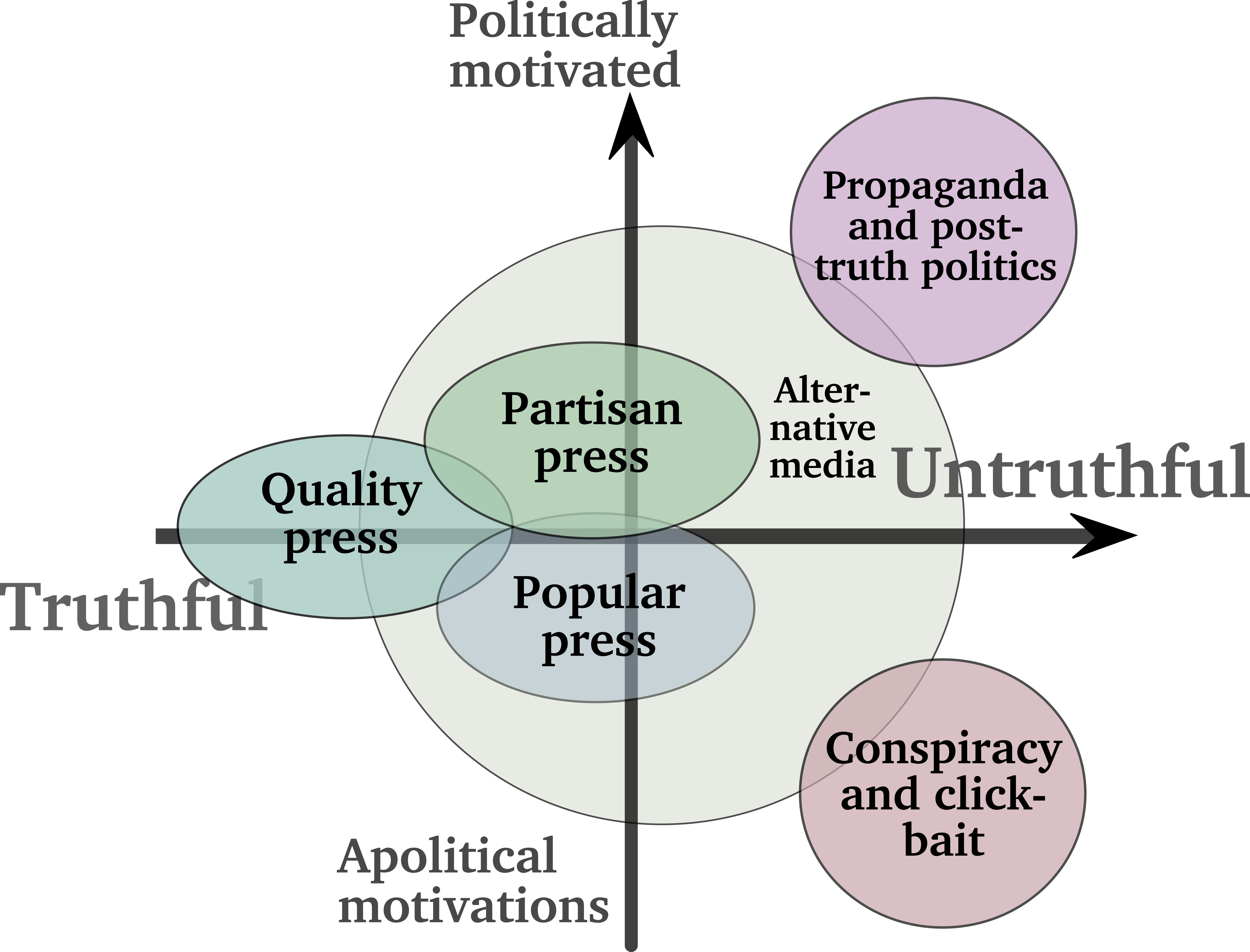}
  \captionof{figure}{Term Definition and Conceptualisation}
   \label{fig:defterms}
\end{figure}

The politically-motivated actors and strategies that are central to this
study (partisan media, propaganda, and post-truth politics) have complex, overlapping characteristics. Figure~\ref{fig:defterms} provides a conceptual diagram of these inter-relationships, as examined in the scope of this paper. We distinguish explicitly \textit{political vs. apolitical}, because although there are many other cases where propaganda and partisan media play a significant role, the focus here is on political influences. The sector of the figure that we are interested in in this work is the \textit{top right}; namely, political and less truthful/unbiased, as we aim to highlight these important new trends in techno-political sociology. Others\footnote{\small{\url{https://medium.com/1st-draft/fake-news-its-complicated-d0f773766c79}}} have explored the ``Ps'' concept with more coverage of apolitical motivations.

Inevitably there is overlap and grey areas between the media and
behaviours we discuss in this work. Motivations for behaviours are
unclear; for example, is a popular political message in the press
intended to influence political outcomes or sell more newspapers? In
this work we confine our interest to media behaviour that is
politically engaged \textit{and} misleading. We therefore define:

\begin{itemize}

\item{\textbf{Partisan media} to to be media presenting themselves as
   news, including:}

\begin{itemize}

   \item{\textbf{Partisan press}; mainstream media unambiguously identifying as providers of news reportage, but who may present partisan materials as
     more factual than they really are;}

   \item{\textbf{Alternative media}; a broad and varied ecosystem of
     new publishers presenting themselves as news, some of whom are politically partisan and therefore of interest;}

     \end{itemize}

   \item{\textbf{Propaganda} to be politically motivated behaviours and
     materials with a primary purpose of influencing toward a particular point of view, see e.g. OED.\footnote{\small{\url{http://www.oed.com/view/Entry/152605}}} Origin may be veiled;}

\item{\textbf{Post-truth politics} to be politically motivated output
   with little regard for truth and public, political figure or
     entity as instigator;}

\end{itemize}

We explore our findings below under these headings.

\vspace{0.25in}
\vspace{0.15in}

\section{Methodology}

The first corpus used is a large collection of tweets collected using the GATE Cloud Twitter
Collector~\footnote{\small{\url{https://cloud.gate.ac.uk/shopfront/displayItem/twitter-collector}}}, a tool that allows tweets to be gathered according to search criteria as they appear, and processed using
GATE~\footnote{\small{\url{https://gate.ac.uk/}}} text processing
pipelines to enrich the tweets with relevant background information,
including the EU membership stance of the author. The method is described more fully by Maynard \textit{et  al}~\shortcite{maynard2017framework}.
In the next section we describe collecting the tweets, then after that the user vote intent classification. The corpus thus enriched was indexed using the M\'imir
search engine for efficient exploration, which again is described in
more detail by Maynard \textit{et  al}~\shortcite{maynard2017framework}.

The second corpus is Twitter's IRA data downloaded from their site.\footnote{\small{\url{https://about.twitter.com/en_us/values/elections-integrity.html#data}}} We introduce this corpus at the end of this section and describe how we classified the accounts into six types introduced by Linvill and Warren~\shortcite{linvill2018troll}. These account types are distinctively different in behaviour, and classifying them enabled important insights.

\subsubsection*{Partisanship Attention Score}

Throughout the work we make use of Partisanship Attention Score (PAS), first introduced by Faris \textit{et
  al}~\shortcite{faris2017partisanship}. This metric is a simple ratio of
the number of times a source is linked by one valence of user, for
example leavers, versus the other valence. In this work we use
``leave-PAS'' to describe a PAS in which leave linkers outnumber
remain linkers, and ``remain-PAS'' to describe a PAS in which remain
linkers dominate. We have grouped sources into five sets; those in
which a PAS is greater than 30:1 (one leave set and one remain set),
those in which the PAS is greater than 3:1 (leave and remain) and
those with a more balanced PAS of less than 3:1. The 30:1 and 3:1
ratios were selected heuristically--throughout the work we are careful
to reflect on how that choice might affect the results.

\subsection{Brexit Tweet Collection}

Around 17.5 million tweets were collected up to and including 23 June
2016 (EU referendum day). The highest volume was 2 million tweets on
Jun 23rd (only 3,300 lost due to rate limiting), with just over 1.5
million during poll opening times. Of the 2 million, 57\% were
retweets and 5\% replies. June 22nd was second highest, with 1.3
million tweets. The 17.5 million tweets were authored by just over 2
million distinct Twitter users (2,016,896). The work presented here
focuses on a subset of these, covering the month up to and including
June 23rd. Within that period, there were just over 13.2 million
tweets, from which 4.5 million were original tweets (4,594,948), 7.7
million were retweets (7,767,726) and 850 thousand were replies
(858,492). These were sent by just over 1.8 million distinct
users. The tweets were collected based on the following keywords and
hashtags: \textit{votein, yestoeu, leaveeu, beleave, EU referendum,
  voteremain, bremain, no2eu, betteroffout, strongerin, euref,
  betteroffin, eureferendum, yes2eu, voteleave, voteout, notoeu,
  eureform, ukineu, britainout, brexit, leadnotleave}. These were
chosen for being the main hashtags, and are broadly balanced
across remain and leave hashtags, though the ultimate test of the
balance of the dataset lies in the number of leavers and remainers
found in it, which is discussed below.

Most URLs found in tweets have been shortened, either automatically by
Twitter or manually by the user, which has the side effect of
obfuscating the original domain being linked to. For this work we
expanded the URLs in tweets using the following approach. From manual
analysis of the URLs we accumulated a list of 18 URL shorteners or
redirect services: shr.gs, bit.ly, j.mp, ow.ly, trib.al, tinyurl.com,
ift.tt, ln.is, dlvr.it, t.co, feeds.feedburner.com,
redirect.viglink.com, feedproxy.google.com, news.google.com,
www.bing.com, linkis.com, goo.gl, and adf.ly. All URLs from other
domains were considered to already be expanded. (A small number of
minor URL shorteners have gone unexpanded due to the long tail in this
large tweet set and the necessity of manually identifying shortening
services.) When we saw a shortened URL it was expanded, either by
following HTTP redirects or using the API of the shortener,
recursively until the resulting URL no longer pointed to a domain in
our list of shorteners.

\subsection{User Vote Intent Classification}

Classification of users according to vote intent was done on the basis
of tweets authored by them and identified as being in favour of
leaving or remaining in the EU. Such tweets were identified using 59
hashtags indicating allegiance, given in the online experimental
materials\footnote{\small{\url{http://http://staffwww.dcs.shef.ac.uk/people/G.Gorrell/publications-materials/brexit-domains-shared-materials.ods}}}
Hashtags in the final position more reliably summarise the tweeter's position, so only these were used. Consider, for example. ``is Britain really \#strongerin? I don't think so!  \#voteleave''.


This approach was evaluated using a set of users that explicitly
declared their vote intent. A company called
Brndstr\footnote{\small{\url{http://www.brndstr.com/}}} ran a campaign
offering a topical profile image modification (a flag overlaid on their profile picture) 
in response to a formulaic vote intent declaration mentioning their brand. This enabled a ground truth sample to be easily and accurately gathered. On these data, we found our method produced a 94\% accuracy even on the basis of a single partisan tweet (where three are required, an accuracy of 99\% can be obtained, though only 60,000 such users can be found, as opposed to 290,000 with at least one partisan tweet). The Brndstr data itself, consisting of around 100,000 users of each valence, was also used to supplement the set, raising the accuracy further, and resulting in a list of 208,113 leave voters and 270,246 remain voters. Table~\ref{tab:votercl} gives detailed statistics for three conditions; one matching tweet found for that user, two found or three found. ``Total'' is the total number of users found with that number of matching tweets. ``Brndstr found'' is the number of those users found in the Brndstr set, and so able to be evaluated. The remaining figures refer to that set, providing an accuracy for the total list of users found using the given minimum number of partisan tweets.

\begin{table}
\begin{center}
\resizebox{\columnwidth}{!}{%
\begin{tabular}{|l|c|c|c|c|c|}
\hline
 & \textbf{Total} & \textbf{Brndstr} & \textbf{Of found} & \textbf{Accuracy} & \textbf{Cohen's}\\
 & & \textbf{found} & \textbf{correct} & & \textbf{kappa} \\
\hline
Leavers, 3\# & 34539 & 1142 & 1129 & 0.987 & 0.972\\
Remainers, 3\# & 26674 & 603 & 594 & &\\
\hline
Leavers, 2\# & 49080 & 1368 & 1350 & 0.984 & 0.966\\
Remainers, 2\# & 50972 & 901 & 882 & &\\
\hline
Leavers, 1\# & 114519 & 1935 & 1801 & 0.943 & 0.885\\
Remainers, 1\# & 175042 & 1744 & 1667 & &\\
\hline
\end{tabular}
}
\end{center}
\captionof{table}{Brexit Classifier Accuracy}
\label{tab:votercl}
\end{table}
\vspace{0.2in}

There may be a case for using a threshold of two hashtags in order to
produce a more balanced set of leavers and remainers, but this would
disproportionately exclude remainers with more moderate feelings (if
the number of hashtags can be seen as an indicator of this). The
resulting set is somewhat slanted toward remainers, demonstrating the
obvious; that Twitter isn't a representative sample of the UK
population, who voted to leave the EU to the order of 52\%. However,
leavers were more vocal and apparent in the data presented below,
contrary to what we would expect if the higher number of remainers had
affected the result. It is possible that some users changed their mind
about how to vote after making their Brndstr declaration, but voters
making an online declaration of their vote intent are perhaps those
less likely to vacillate, and the work can in either case be seen as
an exploration of the behaviour of those who held a particular
allegiance during the time period studied.

\subsection{IRA Corpus and Account Classification}\label{sec:ira-dataset}

The Twitter IRA corpus\footnote{\small{\url{https://about.twitter.com/en_us/values/elections-integrity.html#data}}} contains 3,836 unique users and 9,041,308 tweets. The tweets are posted in 57 different languages, but most of the tweets are in Russian (53.68\%) and English (36.08\%), comprising almost 90\% of the tweets. The majority of accounts (as opposed to tweets) are self-declared English language (2,384), but note that many of these have Russian display names. Average account age is around four years, and the longest accounts are as much as ten years old. Linvill and Warren~\shortcite{linvill2018troll} have analyzed the English language accounts and find several key types of account emerging. A large amount of activity in both the English and Russian accounts is given to \textbf{news} provision. Secondly, many accounts seem to engage in \textbf{hashtag games}, which may be an easy way to establish a history for an account to make it seem more credible. Of particular interest however are the political trolls. \textbf{Left trolls} pose as individuals interested in the Black Lives Matter campaign. \textbf{Right trolls} are patriotic, anti-immigration Trump supporters. Among left and right trolls, several have achieved large follower numbers and even a degree of fame.\footnote{\small{\url{https://www.theguardian.com/technology/shortcuts/2017/nov/03/jenna-abrams-the-trump-loving-twitter-star-who-never-really-existed}}} Finally there are \textbf{fearmonger} trolls, that propagate scares, and a small number of \textbf{commercial} trolls. The Russian language accounts may also provide news, or may pose as individuals with opinions about for example Ukraine or western politics. These troll types provide insight into how IRA effort was targeted and to what extent these different behaviour types translate into impact, such as followers attracted to the accounts and retweets achieved. For this reason we took their dataset and built a classifier enabling us to classify all the accounts.

Linvill and Warren manually categorized 1,102 IRA-associated handles into the six categories described above, providing us with an adequate training set to build a classifier. 55\% of their labelled accounts are right trolls, 20\% are left trolls, 10\% are fearmonger and hashtag gamer accounts, 5\% are newsfeeds and less than 1\% are commercial accounts. We used a support vector machine (SVM) to predict the categories of the remaining accounts. Features were term frequency-inverse document frequency (tf-idf) of English tweet texts, the domain of shared links including the domains of the shortened and expanded versions of the links, and the topic distribution of the tweet text.

We used 75\% of the dataset for training and 25\% for testing. The F1 score was 0.89, which was also equal to precision and recall values. The final model was trained on all data and was used to classify the remaining 2157 accounts which had English tweets. No attempt was made to classify an account that had no English language tweets. The resulting fully classified dataset contains 60\% right trolls, 12\% fearmongers, 11\% having no English language tweets, 10\% left trolls, 5\% hashtag gamers, 3\% newsfeed accounts and negligible commercial accounts (n=6). The reason for the change in class proportions is likely to be the criteria that Linvill and Warren used for selecting accounts to manually classify. They classified accounts represented in their tweet set, which was collected via retrospective search on IRA account names in late 2017, and collected therefore only tweets still available at that point going back to mid-2015. We find generally speaking more left and right trolls than in their sample, and less newsfeeds and hashtag gamers.

\section{Findings}

We now present findings under the headings of the three ``Ps'', beginning with partisan media, then moving on to propaganda, then post-truth politics.

\subsection{Partisan Media}
\label{sec:partisan}

\begin{figure}
 \centering
  \includegraphics[width=0.40\textwidth]{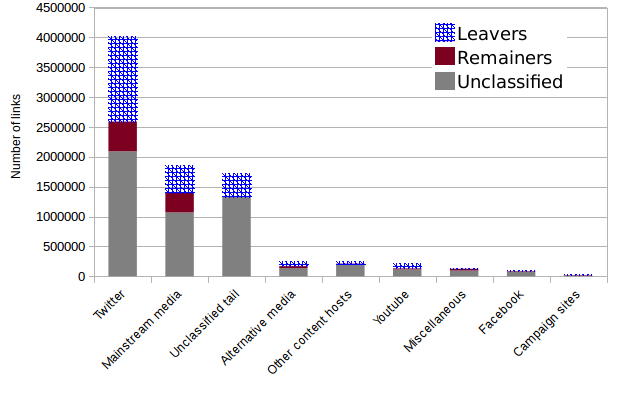}
  \captionof{figure}{Types of links posted}
  \label{fig:typelink}
 \end{figure}

We begin our investigation with the Brexit tweet collection described above. As a starting point for quantifying the various influences and evidence of partisanship, the top 100 most posted domains were manually grouped into high level categories, as shown in figure~\ref{fig:typelink}. The dominant domain to appear was Twitter itself, appearing whenever anyone posts an image, as well as when they link to another tweet. After that, the greater proportion of the links are to items in a wide variety of mainstream news media. ``Other content hosts'' refers to smaller content platforms such as Instagram. YouTube and Facebook are listed separately. Finally, smaller amounts of material are linked from referendum campaign sites and alternative media. (Alternative media range from publications that are nearly mainstream through to conspiracy sites and fake news.) The ``long tail'' of a further 17,000 less linked domains that haven't been
manually classified are included in the chart to give a quantification
of the unknown; note that this unknown section is likely to contain
many more small alternative media, blogs etc. than mainstream
media. Also only domains that were tweeted at least once by a user
that has been classified for vote intent were included. The actual
number of domains mentioned in the set is much greater. The graph broadly agrees with table 1 of Narayanan \textit{et
  al}~\shortcite{narayananrussian}. We are also able divide each count into
three parts, indicating the proportion of tweets in that section by
unclassified users, remainers and leavers. It is evident at a glance
that remainers were tweeting less linked material, since their
representation is smaller. Also there were fewer remainers in the
unclassified tail (that is, the column of unclassified sites, not the
unclassified users), suggesting perhaps a preference for more popular
sites on the part of remainers. It is unknown how many leavers,
remainers and undecideds constitute the unclassified users (the grey
bottom section of the columns) but there's no particular reason why
the classified users wouldn't give a representative impression.

\vspace{0.15in}
\vspace{0.15in}

\subsubsection*{PAS of High Impact Media}

\begin{figure}
 \centering
  \includegraphics[width=0.5\textwidth]{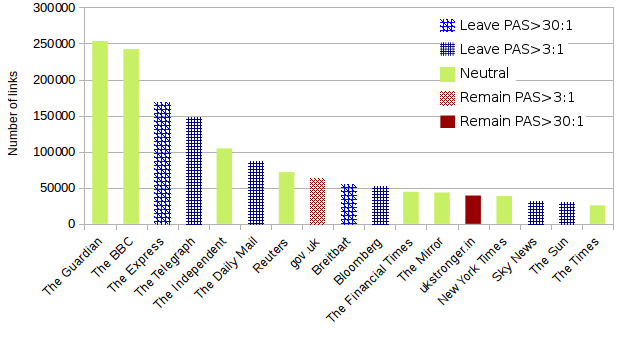}
  \captionof{figure}{Number of appearances of high impact sites}
  \label{fig:biggestsites}
 \end{figure}

Figure~\ref{fig:biggestsites} shows the sites that had the most
impact, in terms of total number of times they appeared in tweets in
the Brexit dataset. These were almost entirely mainstream media,
mostly UK media, with the exception of the remain campaign site
``ukstronger.in'' and the UK government domain. The graph gives total
counts of appearances of the most influential domains, colour coded by
partisanship attention score (PAS); the ratio of links from leave
voters to remain voters or vice versa. Platforms such as Facebook,
where the site doesn't author the content, are excluded. Only link
appearances in original tweets are used in this graph (not appearances
in retweets or replies). Tables~\ref{tab:neutralishsites}
and~\ref{tab:partisansites} in the appendix give a longer
list of sites. The full set is also available for
download~\footnote{\small{\url{http://http://staffwww.dcs.shef.ac.uk/people/G.Gorrell/publications-materials/brexit-domains-shared-materials.ods}}}

\vspace{0.15in}
\vspace{0.15in}

On page 13 of Moore and Ramsay~\shortcite{moore2017uk} a similar graph
shows the number of referendum-related articles published by UK
media. The number of Brexit articles published by a medium shows a
strong correlation to its link presence on Twitter (0.71). In fact,
the Express has been somewhat less taken up on Twitter than its
engagement with the subject might predict; figure~\ref{fig:grassroots}
and its discussion later in the paper may offer further insights on
this point.

It is evident that mainstream media were the dominant source of linked
materials in the Brexit discussion on Twitter, with the six most
influential domains all being British mainstream media as shown in
figure~\ref{fig:biggestsites}. Smaller in influence but nonetheless
significant were alternative media, with Breitbart appearing in ninth
place in figure~\ref{fig:biggestsites}, user-shared content on other
content platforms such as Facebook, and campaign sites. This suggests
a continuing important role for traditional media, though leaves
questions about how social media, and
indeed alternative media, may interact to popularize certain materials
and influence the focus. It is also apparent that the most popular
domains were either neutral in their appeal or appealed to leavers, with
only two smaller sources, the government and the ``Stronger In''
campaign, appealing to remainers. This subject is taken up more fully
in the next section.

\subsubsection{Ground-Truthing Mainstream Media}

\begin{figure}
  \centering
  \includegraphics[width=.45\textwidth]{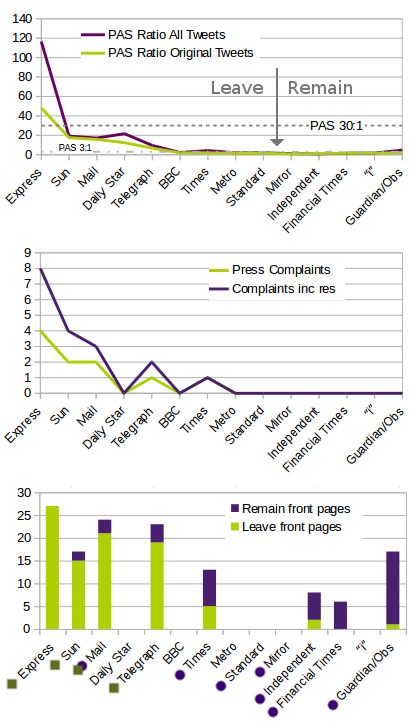}
\captionof{figure}{PAS (a), Press Complaints (b) and Partisan Front Page Counts (c) for UK Mainstream News Media}
\label{fig:mediabias}
 \end{figure}

Figure~\ref{fig:mediabias}a shows British mainstream newspapers ranked from left to right in order of their PAS ratio. For those media with negative leave PAS ratios, the remain PAS ratio has been plotted (ratio of appearances in remain tweets against those in leave tweets). In this way, both leave and remain media can be shown commensurately on the same graph. The point at which the PAS ratios switch direction is indicated with a vertical arrow. The extreme right of the graph, therefore, shows the newspaper with the highest remain PAS ratio (The Guardian/Observer). Two horizontal lines indicate PAS ratios of 3:1 and 30:1. PAS ratios for link appearances in all tweets and just original tweets are shown.

In figure~\ref{fig:mediabias}b, the green line indicates the number of upheld press complaints for that medium. The purple line also includes the number of complaints for which a resolution was found. The majority of press complaints regarded articles that were anti-immigration in their focus.

In figure~\ref{fig:mediabias}c, newspaper front pages provided by Moore and Ramsay~\shortcite{moore2017uk} for the two month period preceding the referendum have been manually classified as leave, remain or neutral in their orientation. An example of a leave front page might be "EU 'very bad' for pensions" (The Express, June 21st 2016). An example of a remain front page might be "Vote remain today" (The Mirror, June 23rd 2016). Bars show leave front pages in green and remain in purple. Where possible, the original article was consulted before classifying a front page. However, in many cases this information wasn't accessible. In these cases, a conservative judgment was reached, but this means that counts for the Sun and the Independent may be a little depressed, since the full article usually wasn't available for them. Note also that the work was completed by a single annotator, and that in many cases, classifying the headlines was quite a subtle judgment call.

\vspace{0.15in}
\vspace{0.15in}

Several British newspapers declared their allegiances regarding Brexit, reportedly giving media supporting the UK leaving the EU an audience of around 4.8 million, while those in favour of remaining in the EU reach just over 3
million~\footnote{\small{\url{https://www.huffingtonpost.co.uk/entry/which-newspapers-support-brexit_uk_5768fad2e4b0a4f99adc6525}}}. Stance information is included in figure~\ref{fig:mediabias}c in the form of coloured marks--a green square for leave and a purple circle for remain. Both marks appear for the Mail because the Daily Mail shares its domain with the Mail on Sunday. The Daily Mail were in favour of leaving the EU, and the Mail on Sunday, with a slightly lower circulation, were in favour of remaining.

PAS was found to correlate with press complaints (0.922, p\textless0.001) as well as bias as quantified by the magnitude of the difference between pro- and anti-Europe front page counts (0.842, p\textless0.001).

Figure~\ref{fig:mediabias}a shows that all of the media that declared their support for the remain cause were broadly neutral in their appeal, with the exception of the Guardian/Observer, who, when retweets and replies are counted, has a leave PAS greater than 3:1. The media that declared their official support for leave all to varying extent appealed more to leavers. This brings to mind Faris \textit{et al}'s~\cite{faris2017partisanship} conclusion from their study of the 2016 US presidential election that mainstream media ranging from left to centre right show more investment in principles of neutrality. The Brexit question cut across the political spectrum, although in terms of media stance, the left-leaning papers favoured remain and the right, leave. However, it is also possible that leavers engaged with
remain materials for other reasons. Press complaints and front page partisanship data provide further insights. It is interesting to note that PAS seems to echo upheld press complaints better than it does partisanship as indicated by front pages. There are prominent cases where media published many stories in keeping with their Brexit stance, but without attracting press complaints; most notably the Telegraph and the Guardian. Materials supportive of a particular stance don't \textit{per se} seem to draw partisan attention---the PAS of both these media is low.

This is important in correctly interpreting figure~\ref{fig:biggestsites}. The medium with the biggest impact is the Guardian, which published many pro-remain articles. So in this sense, there wasn't a lack of attention to pro-remain materials, and if the colour coding of the graph were based on the ``front page diff'' used above, the impression created would be quite
different. PAS captures something different. Manual review of the tweets suggests that Guardian articles tend to be factual in tone, and attract critical engagement from leavers. Express articles tend to use emotive and suggestive language, and seem to attract less discussion. Moore and Ramsay's analysis~\shortcite{moore2017uk} gives much information about the rhetorical styles employed by the press in the run-up to the referendum. Circulation size does not explain the number of complaints received, with the Express having less than half the readership of any of the four largest
media.\footnote{\small{\url{http://www.pressgazette.co.uk/nrs-national-press-readership-data-telegraph-overtakes-guardian-as-most-read-quality-title-in-printonline/}}}

\subsubsection*{Extreme/Affective Materials}

We saw in section~\ref{sec:partisan} that high PAS scores show a potential relationship with upheld press complaints, and that polarity of PAS is a good indicator of the stance of the source, as determined from press front pages. We now use PAS scores of greater than 30:1 to select sources that may be misleading for further examination. Sites of either camp with at least 1000 total mentions in tweets in the dataset and at least 50 tweets, retweets or replies by leavers or remainers were manually analysed. We present the sites divided into 4 categories; mainstream media, alternative media, campaign sites and other sites. ``Others'' includes for example personal blogs or special interest websites not primarily focused on Brexit.

\vspace{0.15in}
\begin{figure}
 \centering
  \includegraphics[width=0.40\textwidth,height=2.5in]{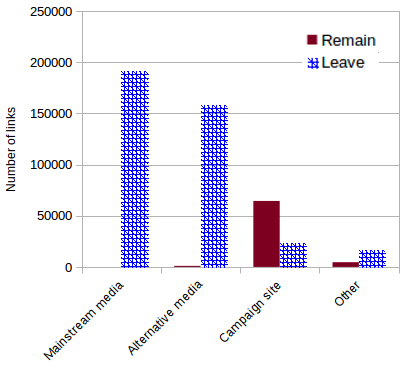}
  \captionof{figure}{Who are the PAS\textgreater30:1 influencers?}
  \label{fig:whohighlypreferred}
 \end{figure}
\vspace{0.15in}

\begin{figure}
 \centering
  \includegraphics[width=0.45\textwidth]{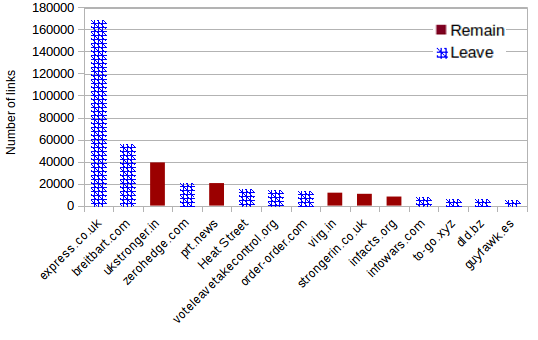}
  \captionof{figure}{Who are the PAS\textgreater30:1 sites?}
  \label{fig:whohp}
\end{figure}

Figure~\ref{fig:whohighlypreferred} shows that remain
PAS\textgreater30:1 sites are dominated by explicit campaign sites. As
we would expect given the data above, among leave influencers we see
more mainstream media---note that the only high PAS mainstream media
were leave media; namely the Express. We also see a much greater role
for alternative media in the leave campaign. The total impact of leave
PAS\textgreater30:1 media was 389,000 mentions. For remain it was
70146 mentions, or 18\% of the PAS\textgreater30:1 impact. All sites
with a PAS higher than 30:1 and more than 5000 mentions are shown in
figure~\ref{fig:whohp}. The Express dominates, with the US alternative
medium Breitbart in second place. As indicated above, remain sites are
mainly campaign sites. Other leave sites are media ranging from
alternative to conspiracy, plus the campaign site
``voteleavetakecontrol.org''. A longer list can be found in
table~\ref{tab:partisansites} in the appendix.

\vspace{0.15in}
\vspace{0.15in}

Key observations from figure~\ref{fig:whohighlypreferred} include that
in terms of mentions in tweets, the influence of leave sites dwarfs
that of remain sites. It is also notable in that figure that high
remain-PAS sites were mostly explicit campaign sites; in other words,
openly partisan, with no suggestion of providing reportage. The range
of media providing high leave-PAS materials, plus the presence of
Breitbart raises the question of whether these findings demonstrate a
similar phenomenon happening in the UK as described by Faris
\textit{et al}, or whether indeed it is simply the same phenomenon -
an extension of the same network of propaganda.

\begin{figure}
  \centering
  \includegraphics[width=\linewidth]{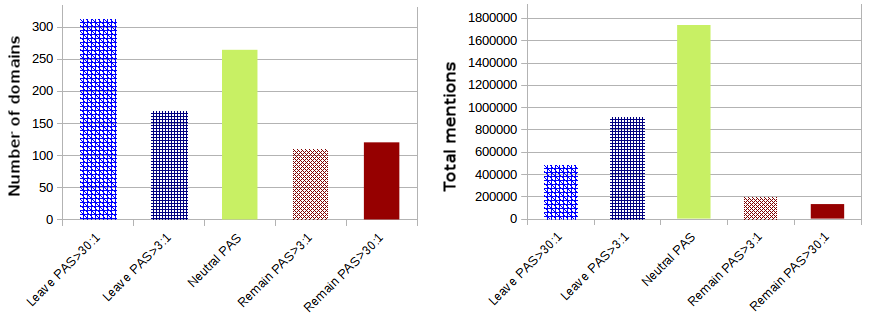}
  \captionof{figure}{All domains vs total mentions by PAS of domain}
  \label{fig:grassroots}
\end{figure}

Figure~\ref{fig:grassroots} presents counts of sites according to
their PAS status. A threshold of 20 total original tweets by leavers
and remainers was applied, in order to exclude sites for which too
little evidence was available to classify them. The graph shows peaks
to either extreme, despite the stringent 30:1 criterion, reinforcing
previous researchers' findings that extreme content tends to
proliferate on social
media~\cite{faris2017partisanship,silverman2015lies,barbera2015understanding,preoctiuc2017beyond}. The
neutral peak most likely arises because content-neutral platforms such
as Facebook are counted here, rather than because there is a peak in
neutral materials such as unbiased news providers. On the right we see
the actual link counts to the sites. Twitter mentions have not been
included, since they give a large, uninformative boost to the neutral
count. Were other content-neutral platforms to be excluded, this count
would be lower still. Nonetheless, we see that the extremes no longer
outnumber the moderate sites. Evidently most Twitter users prefer less
extreme materials of those on offer. However, this provides evidence of the diet Twitter is offering.

\vspace{0.15in}
\vspace{0.15in}

\subsection{Online Propaganda}
\label{sec:propaganda}

\begin{figure*}
  \centering
  \includegraphics[width=.75\textwidth,height=10.5cm]{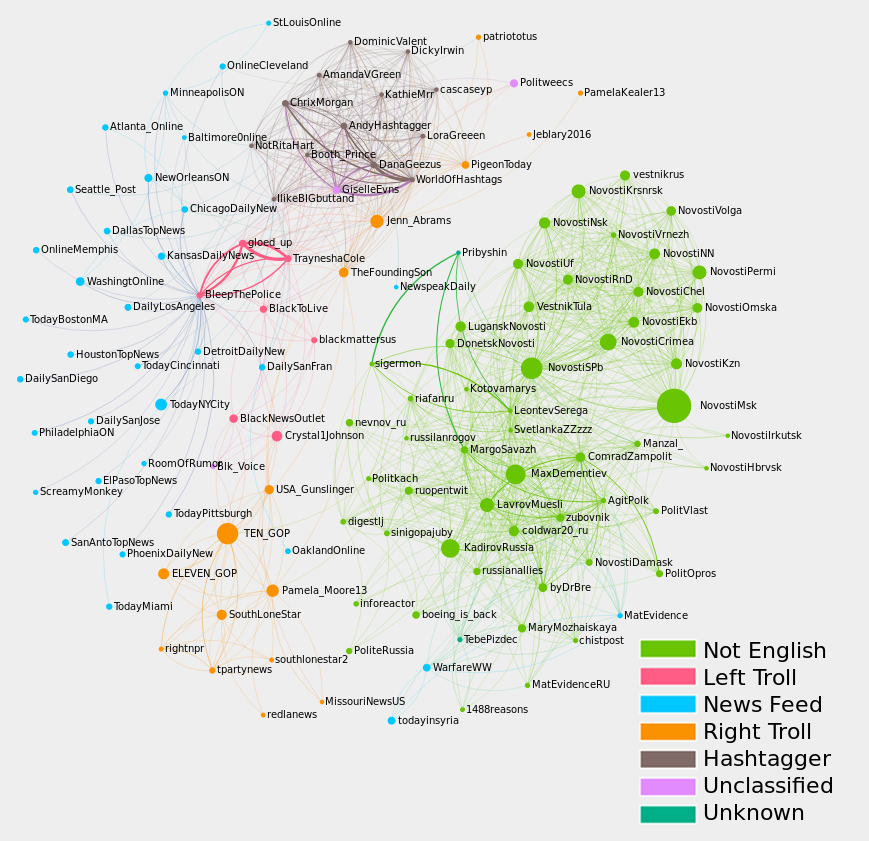}
\captionof{figure}{Network of IRA Troll Accounts}
\label{fig:toptrolls}
 \end{figure*}

Recall that political propaganda is non-objective information, which is aimed at influencing citizens and/or furthering a political agenda. In this section we use the Twitter IRA tweet collection, introduced in Section~\ref{sec:ira-dataset}, to explore evidence for the impact of different propaganda strategies.

\vspace{0.15in}
\begin{figure}
  \centering
  \includegraphics[width=.35\textwidth]{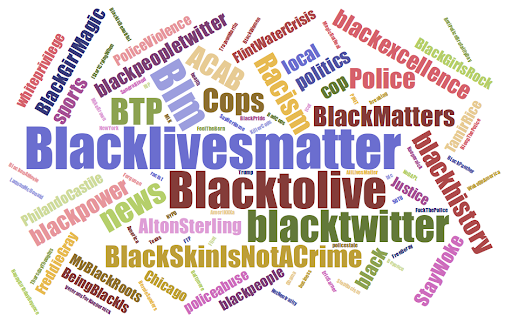}
\captionof{figure}{Left Troll Hashtags}
\label{fig:left-hashtags}
 \end{figure}
\vspace{0.15in}

\vspace{0.15in}
\begin{figure}
  \centering
  \includegraphics[width=.35\textwidth]{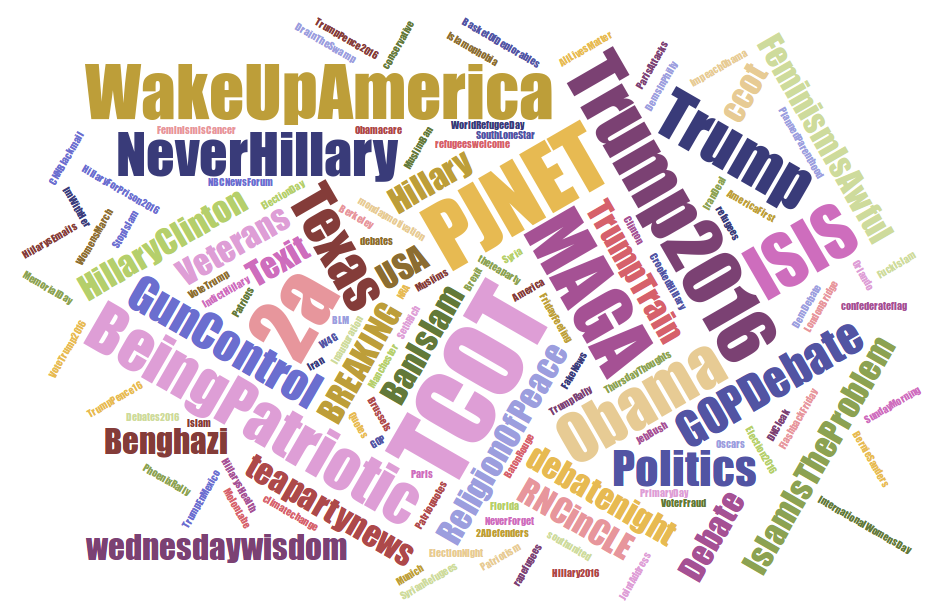}
\captionof{figure}{Right Troll Hashtags}
\label{fig:right-hashtags}
 \end{figure}
\vspace{0.15in}
 
Initially, in the autumn of 2017, Twitter released a list of around 3,000 Twitter accounts to US Congress that they had identified as being Russian state-controlled troll accounts, and had suspended. In the autumn of 2018, the full set of 9 million tweets by these IRA propaganda accounts were released. The majority of tweets are in Russian as noted above, primarily with Ukraine-related focus. In contrast, the English language tweets focus predominantly on US politics.

 
Prior to the release of the full 9 million tweet set, Linvill and Warren~\shortcite{linvill2018troll} researched a partial subset of 3 million  tweets by most of the IRA accounts, which they gathered and released independently. They found differing patterns of troll activity, with news accounts keeping up a relatively steady output of genuine news and achieving a fair reach, hashtag trolls showing bursty activity around playing ``hashtag games''\footnote{\small{\url{https://www.huffingtonpost.com/jeffrey-dwoskin/you-should-be-playing-has_b_7910728.html}}} (i.e., seeking to get many retweets and favourites through exploiting hashtags), and left and right trolls being more event-triggered. Political trolls in some cases achieve a significant following. Examples are given in table~\ref{tab:toptrolls}, and include both left and right trolls and news feeds.

Figures~\ref{fig:left-hashtags} and~\ref{fig:right-hashtags} give word clouds we generated for the subset of left and right troll accounts that were manually identified by Linvill and Warren~\shortcite{linvill2018troll}. Left troll material has a strong Black issues focus, and often talks about conflict with the police. Right troll material is political, supportive of Trump, against the Democrats and anti-Muslim.\footnote{``TCOT'' means ``Top Conservatives on Twitter''; ``PJNET'' means ``Patriot Journalist Network''.} We also find differences in the web domains left and right trolls tend to link. The most-linked domains of we found for Linvill and Warren's left and right trolls are included in table~\ref{tab:irasites} in the appendix. Domains intersect with domains linked by leavers and remainers, as described above and also included in the appendix. Three sites frequently linked by left trolls appear on the Brexit list; the Independent, the Huffington Post and the New York Times. All had a neutral PAS. Three highly hyperpartisan sites frequently linked by right trolls also appear on the Brexit list; Breitbart, Infowars and the Express. All had a leave PAS of greater than 30:1. This suggests an overlap in outlook between Brexit leave voters and the right troll persona. Left trolls link neutral sites as well as Black-focused sites that aren't relevant to Brexit.


\vspace{0.15in}
\begin{table}
\begin{center}
\resizebox{\columnwidth}{!}{%
\begin{tabular}{|l|l|l|l|l|l|l|}
  \hline
  \textbf{Type} & \textbf{Number} & \textbf{Av Tw} & \textbf{Av Orig} & \textbf{Retw Rec} & \textbf{Av Foll} & \textbf{Retw Rat}\\
  \hline
  Right & 2194 & 2560 & 1436 & 8710 & 1609 & 6.066\\
  Left & 339 & 2755 & 1025 & 30121 & 1815 & 29.377\\
  Fearmonger & 432 & 487 & 481 & 10 & 62 & 0.022\\
  Hashtag & 189 & 3041 & 1582 & 924 & 2225 & 0.584\\
  News & 99 & 9981 & 9859 & 13925 & 9552 & 1.412\\
  \hline
  All trolls & 3667 & 2466 & 1537 & 8522 & 1741 & 5.546\\
  \hline
\end{tabular}
}
\end{center}
\captionof{table}{Troll Impact}
\label{tab:impact}
\end{table}
\vspace{0.15in}

Table~\ref{tab:impact} gives impact statistics for the different troll types. First we give average number of tweets, then average number of original tweets (excluding retweets). Then we report average number of retweets received, average number of followers and rate of retweets per original tweet. It is clear that political trolls achieve by far the best ratio of retweets to original tweets. Left trolls achieve more retweets per original tweet than right trolls, both in terms of mean, shown in the table, and median (48 vs. 21). However, other account types are more highly followed, and news and hashtag accounts may influence their followers even though their tweets do not inspire retweets to the same extent. Where an agent retweets someone else's tweet rather than authoring an original tweet, we don't have data about how widely retweeted that tweet was, as it counts for the original author; it is possible that agents retweeting the tweets of others are having significant impact in amplifying a message. Of the account types shown, all have average longevities of active life approaching a couple of years with the exception of fearmonger trolls, where the average duration of active life (first activity to last activity) is less than six months. Follower count correlates with retweet rate per original tweet to the tune of 0.25, which is highly significant, but as we see, different types of tweeting behaviour produce different profiles in terms of being followed and being retweeted.

\vspace{0.15in}
\begin{figure}
  \centering
  \includegraphics[width=\columnwidth]{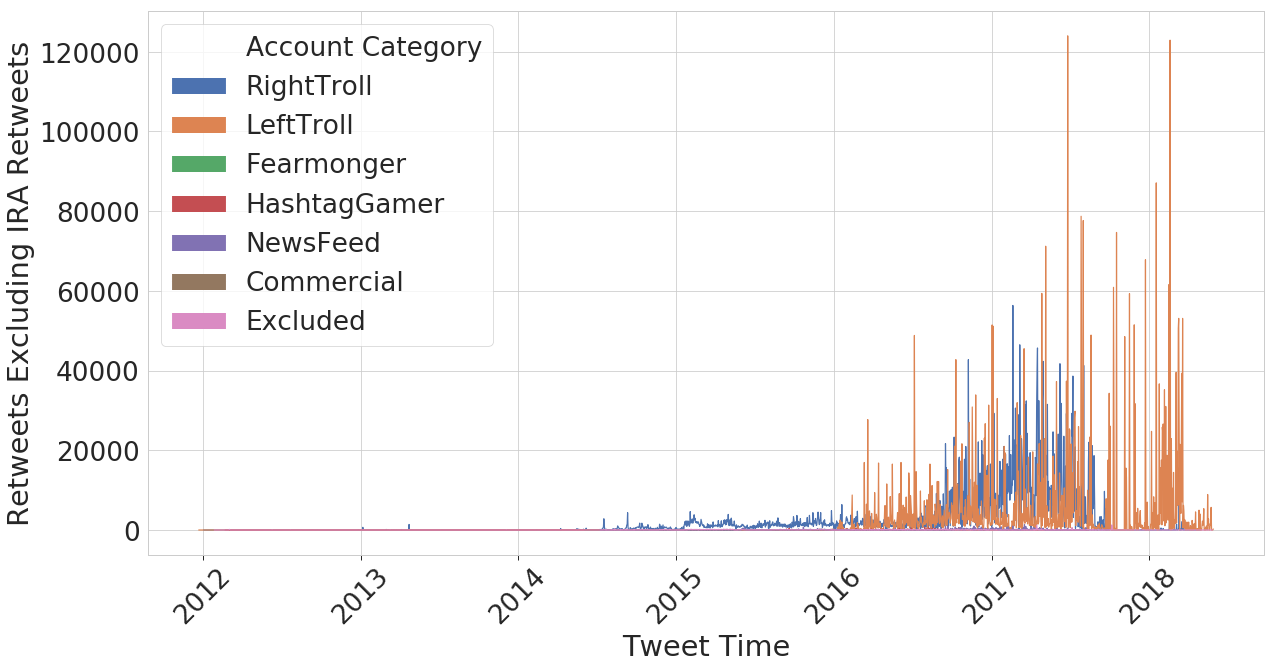}
\captionof{figure}{Timeline of Retweets Achieved by Troll Type}
\label{fig:retweettimeline}
 \end{figure}
\vspace{0.15in}

In figure~\ref{fig:retweettimeline} we see a timeline of retweets achieved for the different types of trolling behaviour. This gives an indicator of the effectiveness of the different troll types. The figure is available in interactive form at \url{https://gate.ac.uk/ira/retweet_counts_excluding_from_trolls.html}. It is notable that political trolls are achieving many more retweets than any other type, with the others barely appearing in the graph. Retweets by other IRA trolls have been removed from these counts. As a whole, IRA trolls have not tended to retweet each other a great deal; 27\% of retweets in the corpus are of other trolls, but this was extremely variable; right trolls retweeted each other significantly until the end of 2016 then stopped. Hashtag gamers do retweet each other to a minor extent.

\begin{table}
\begin{center}
\resizebox{\columnwidth}{!}{%
\begin{tabular}{|l|l|l|l|}
  \hline
  \textbf{Name} & \textbf{Bio} & \textbf{Followers} & \textbf{Tweets}\\
  \hline
  TEN\_GOP & Unofficial Twitter of Tennessee Republicans. & 147,767 & 10,794\\
  &  Covering breaking news, national politics, & &\\
  &  foreign policy and more. \#MAGA \#2A & &\\
  \hline
  Jenn\_Abrams & Calm down, I'm not pro-Trump. I am pro-& 79,152 & 25,378\\
  & common sense. Any offers/ideas/questions? & & \\
  & DM or email me jennnabrams@gmail.com & & \\
  & (Yes, there are 3 Ns) & & \\
  \hline
  Pamela\_Moore13 & Southern. Conservative. Pro God. Anti & 72,121 & 6,203\\
  & Racism & & \\
  \hline
  TodayNYCity & New York City's local news on Twitter. & 66,980 & 59,420\\
  & Breaking news, sports, events and & & \\
  & international news. Tweet us or DM & & \\
  \hline
  ELEVEN\_GOP & This is our back-up account in case & 59,279 & 115\\
  & anything happens to @TEN\_GOP & & \\
  \hline
  wokeluisa & APSA. \#Blackexcellence. Political science & 57,295 & 2,288\\
  & major & & \\
  \hline
  Crystal1Johnson & It is our responsibility to promote the positive & 56,581 & 7,915\\
  & things that happen in our communities. & & \\
  \hline
  SouthLoneStar & Proud TEXAN and AMERICAN patriot \#2a & 53,999 & 3,600\\
  & \#prolife \#Trump2016 \#TrumpPence16  Fuck & & \\
  & Islam and PC. Don't mess with Texas! & & \\
  \hline
\end{tabular}
}
\end{center}
\captionof{table}{High Impact IRA Trolls}
\label{tab:toptrolls}
\end{table}
\vspace{0.15in}

Figure~\ref{fig:toptrolls} gives a network diagram of only trolls with more than 5,000 followers. Connections are based on the trolls mentioning, retweeting, replying to or quoting each other, not whether they follow each other, as we do not have access to that information in the dataset released by Twitter. ``Not English'' accounts are mostly Russian, and consist of a large number of newsfeed accounts (``novosti'') as well as others. The figure is available in interactive form at \url{https://gate.ac.uk/ira/network/}.

In the following subsections we discuss a selection of cases illustrating different aspects of the dataset that shed light on some aspect of online propaganda. We discuss prominent ``spikes''; brief periods of much escalated tweeting. We also briefly cover an attempt at a ``scare'' from 2014, before concluding with an analysis of the relevance of Russian Twitter propaganda to Brexit.

\vspace{0.15in}
\begin{figure}
  \centering
  \includegraphics[width=.5\textwidth]{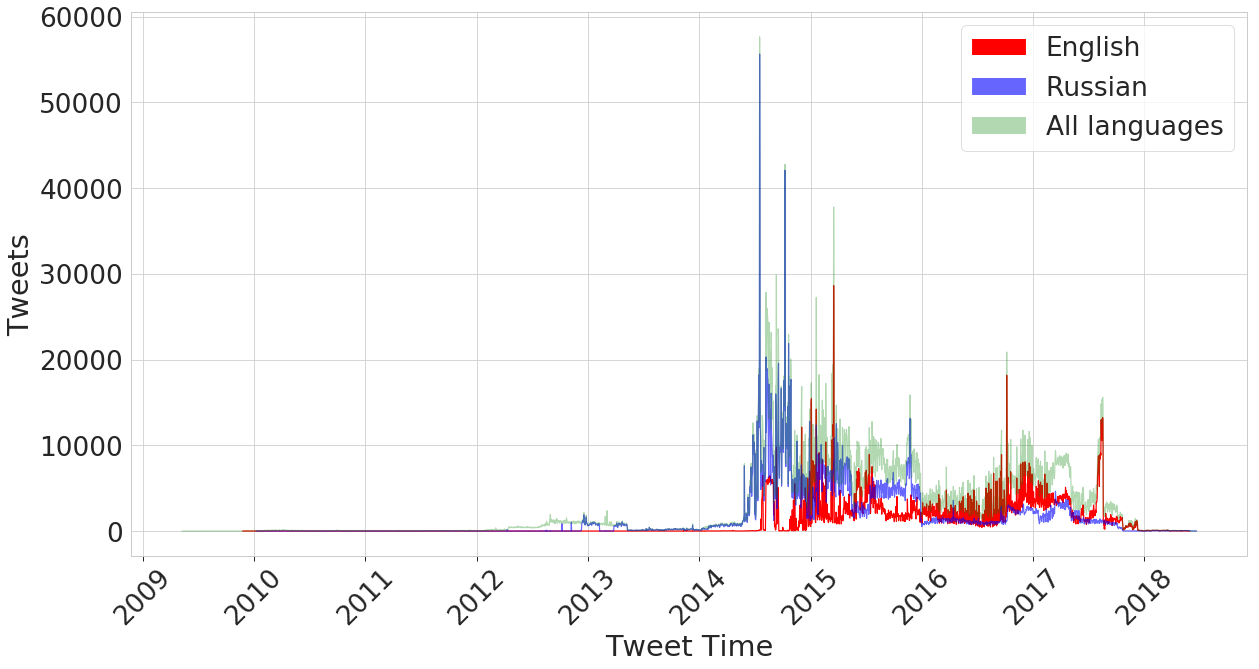}
\captionof{figure}{Timeline of Tweet Activity}
\label{fig:tweet-timeline}
 \end{figure}
\vspace{0.15in}

\subsubsection*{Cases}

There are three prominent spikes in activity among English language tweets, and three among the Russian ones, as can be seen in figure~\ref{fig:tweet-timeline}. The figure is available in interactive form at \url{https://gate.ac.uk/ira/daily_tweets.html}. The first and greatest of the English spikes shows little in the way of meaningful content. Impact (retweets) in this period was negligible despite a high number of original tweets. The second was timed well, in October 2016, as an attempt to influence Americans who would go to the polls to elect a new president the following month. The final of three spikes in English language tweets occurred in August 2017 and focuses on the incidents in Charlottesville.\footnote{\small{\url{https://en.wikipedia.org/wiki/Unite_the_Right_rally}}} Table~\ref{tab:spikes} gives an overview of the spikes. ``\% Retw'' gives the percentage of the tweets that were retweets of others, whereas ``Retw Rec'' gives number of times these tweets were retweeted, and ``Retw Rat'' gives number of times each tweet was retweeted.

\vspace{0.15in}
\begin{table}
\begin{center}
\resizebox{\columnwidth}{!}{%
\begin{tabular}{|l|l|l|l|l|l|}
  \hline
  \textbf{When} & \textbf{Lang} & \textbf{Tweet Total} & \textbf{\% Retw} & \textbf{Retw Rec} & \textbf{Retw Rat}\\
  \hline
  17-20 Jun 2014 & Rus & 118,219 & 17\% & 30,365 & 0.31\\
  8-10 Oct 2014 & Rus & 70,233 & 44\% & 22,572 & 0.57\\
  17-19 Mar 2015 & Eng & 57,710 & 1\% & 637 & 0.01\\
  23-25 Nov 2015 & Rus & 28,252 & 72\% & 38,760 & 4.90\\
  5-7 Oct 2016 & Eng & 31,111 & 90\% & 121,117 & 38.93\\
  11-18 Aug 2017 & Eng & 95,112 & 2\% & 297,960 & 3.20\\
  \hline
\end{tabular}
}
\end{center}
\captionof{table}{Statistics of Tweet Spikes}
\label{tab:spikes}
\end{table}
\vspace{0.15in}

31,111 tweets were found in the set between October 5th and 7th 2016, which constituted the second largest English language ``spike'' in the dataset. It is evident in the table that the number of these tweets that were retweets was high, and at 90\% much higher than the corpus-wide average rate of 38\%.
In the two day window from October 5th to 7th, almost half the tweets originated in the most active twenty accounts and consisted almost entirely of retweets. These accounts had on average 1,300 followers each. Prominent trolls continued their activity as usual during this period, and the top 15, which each had more than 500 retweets and are familiar, established accounts such as ``TEN\_GOP'' and ``Crystal1Johnson'', achieved 98\% of retweets (of original tweets) in this period. The retweet rate of original tweets in this period was 39 retweets per original tweet, which is much higher than the corpus-wide retweet rate of 3.46 retweets per original tweet. It is possible that the retweeting activity boosted the impact of the original tweets during this time; however the retweet quality is generally low and the retweets were not generally of other troll accounts. It is perhaps more likely that the political climate in this period enabled skilled political trolls to be particularly effective.

In the Charlottesville spike we again see the overwhelming majority of retweets achieved by a handful of prominent trolls. 97\% of retweets were achieved by the 19 trolls with retweet counts over 500. Among those 19 we see familiar faces, who continued to operate as usual and with their usual high impact, most notably ``TEN\_GOP'' who achieved 130,000 retweets in that period. However there is also a presence of a cluster of accounts that became active at the end of July 2017 and remained active for short durations only, often posing as patriotic, Trump-supporting individuals and notably giving as their profile URL a link to ``ReportSecret.com'', a now-defunct alternative news site also run by the Internet Research Agency. During the Charlottesville period, one of these accounts achieved 21,000 retweets, a return of four retweets per tweet, particularly notable given that the account was active for only eighteen days. The tone of the material is pro-Trump, consisting of a fair percentage of skilled original tweets, and retweets that are consistent with the message so likely to be manual. 65\% of tweets in this period originated in accounts with ``ReportSecret.com'' profile URLs. This operation was probably more expensive, and whilst the retweet per original tweet rate was typical for the corpus, by far the highest total number of retweets was achieved of all the spikes, whilst succeeding in scaling up beyond the handful of established, popular trolls and gaining significant impact with new accounts.

In contrast, a tweet set from a single day in September 2014 illustrates perhaps a further early unsuccessful attempt at influence. 8,520 tweets in total contained the hashtag ``\#ColumbianChemicals'' and spread false rumours of an accident at a US chemical plant, and consisted of 275 tweets in Russian, most of which came earliest in the day, 3,119 tweets targeted at prominent individuals that achieved just eight retweets, 3,821 original tweets that achieved 1360 retweets, and 1305 retweets by the IRA trolls themselves, accounting for most of the retweets of original tweets. This attempt at a scare clearly fell flat. Here is an example tweet from the set:

\begin{quote}
@BarackObama Barack , Are you kidding?? I saw the video \#ColumbianChemicals and it looks like hell!!! What a nightmare!
\end{quote}

\subsubsection*{IRA and Brexit}

With regards to Brexit, we looked at tweets posted by the accounts in our own Brexit tweet dataset in a one month period before the referendum. Furthermore, using our data, a further forty-five troll accounts were able to be identified and subsequently suspended by Twitter, in work described by Buzzfeed News.\footnote{\small{\url{https://www.buzzfeed.com/tomphillips/we-found-45-suspected-bot-accounts-sharing-pro-trump-pro}}} 
Influence by those accounts was modest. Amongst the 3,200 total tweets, 830 came from the 45 newly identified accounts (26\%). Brexit interest in the new corpus echoed previous findings provided in the Buzzfeed article 
showing little interest in advance of the referendum and a peak on the day of the referendum almost entirely in foreign languages (German).




Table~\ref{tab:hashtags} shows all tweets posted one month before 23 June 2016, which were either authored by Russia Today or Sputnik, or are retweets of these. This gives an indication of how much activity and engagement there was around these accounts. To put these numbers in context, the table also includes the equivalent statistics for the two main pro-leave and pro-remain Twitter accounts. It is evident that influence was modest.

\vspace{0.15in}
\begin{table}
\begin{center}
\resizebox{\columnwidth}{!}{%
\begin{tabular}{|l|l|l|l|l|l|}
  \hline
  \textbf{Account} & \textbf{Orig. tweets} & \textbf{Retweeted}
  & \textbf{Retweets} & \textbf{Replies} & \textbf{Total}\\
  \hline
  @RT\_com & 39 & 2,080 & 62 & 0 & 2,181\\
@RTUKnews & 78 & 2,547 & 28 & 1 & 2,654\\
@SputnikInt & 148 & 1,810 & 3 & 2 & 1,963\\
@SputnikNewsUK & 87 & 206 & 8 & 4 & 305\\
  \hline
\textbf{TOTAL} & 352 & 6,643 & 101 & 7 & 7,103\\
  \hline
@Vote\_leave & 2,313 & 231,243 & 1,399 & 11 & 234,966\\
@StrongerIn & 2,462 & 132,201 & 910 & 7 & 135,580\\
\hline
\end{tabular}
}
\end{center}
\captionof{table}{Russian Account Activity vs Campaign Sites}
\label{tab:hashtags}
\end{table}
\vspace{0.15in}



\subsubsection*{Automation in the Brexit Tweets}

Automation is another area of concern with regards to propaganda, as it may be used to increase reach at low cost. We saw evidence above suggesting that it is difficult to achieve a high impact with automated accounts. However, other research finds a role for automated accounts in information spread~\cite{shao2018spread}. With regards to Brexit, whilst it is hard to quantify automation among the accounts, Bastos and Mircea~\shortcite{bastos2017brexit} identified 13,493
suspected bot accounts, among which Twitter found only 1\% to be
linked to Russia. In our Brexit dataset there are tweets by
1,808,031 users in total, which makes these bot accounts only
0.74\% of the total. If we consider Twitter accounts that have posted more than 50 times a day (widely considered to indicate a high degree of automation), then there are only 457 such users in the month leading up to the referendum on 3 June 2016. The most prolific accounts were "ivoteleave" and "ivotestay", both suspended, which were similar in usage pattern. There were also a lot of accounts that did not really seem to post much about Brexit but were perhaps using the hashtags in order to gain attention for commercial reasons. We also analysed the leaning of these 457 high automation accounts and identified 361 as pro-leave (with 1,048,919 tweets), 39 pro-remain (156,331 tweets), and the remaining 57 as undecided. This leaning towards leave echoes our above findings that the leave campaign was much more vocal on Twitter.

\subsection{Post-Truth Politics--A Tale of Two Claims}
\label{sec:posttruth}

The rise of post-truth politics has been linked to the lowered bar to publication offered by Web 2.0 and the consequent momentum that can be gained for organized disinformation campaigns~\cite{faris2017partisanship}. A House of Commons Treasury Committee Report published on May 2016, states that: ``The public debate is being poorly served by inconsistent, unqualified and, in some cases, misleading claims and counter-claims. Members of both the `leave' and `remain' camps are making such claims.'' We analysed the number of Twitter posts around some of the these disputed claims. A study of the news coverage of the EU Referendum campaign established that the economy was the most covered issue, and in particular, the remain claim that Brexit would cost households £4,300 per year by 2030 and the leave campaign’s claim that the EU cost the UK £350 million each week. Therefore, we focused on these two key claims and analysed tweets about them.

With respect to the disputed £4,300
claim\footnote{\small{\url{https://www.kcl.ac.uk/sspp/policy-institute/CMCP/UK-media-coverage-of-the-2016-EU-Referendum-campaign.pdf}}}
(made by the Chancellor of the Exchequer), we identified 2,404 posts
in our dataset (tweets, retweets, replies), referring to this claim. For the £350 million a week disputed
claim\footnote{\small{\url{https://www.kcl.ac.uk/sspp/policy-institute/CMCP/UK-media-coverage-of-the-2016-EU-Referendum-campaign.pdf}}}
there are 32,755 pre-referendum posts (tweets, retweets, replies) in
our dataset. This is 4.6 times the 7,103 posts related to Russia Today
and Sputnik and 10.2 times more than the 3,200 tweets by the
Russia-linked accounts suspended by Twitter.

In particular, there are more than 1,500 tweets from different voters \textit{within our sample}, with one of these wordings:

\begin{quote}I am with @Vote\_leave because we should stop sending £350
  million per week to Brussels, and spend our money on our NHS
  instead.\\
  \\
  I just voted to leave the EU by postal vote! Stop sending our
  tax money to Europe, spend it on the NHS instead! \#VoteLeave
  \#EUreferendum
  \end{quote}

Many of those tweets have themselves received over a hundred likes and retweets each. This false claim is popularly regarded as one of the key ones behind the success of the leave campaign. Regarding the impact of these claims, a potentially useful indicator comes from an Ipsos Mori poll published on 22nd June 2016, which showed that for 9\% of
respondents the NHS was the most important issue in the campaign.

The leave claim notably appeared as a bus advert, so spreading its message to the voting public via a different channel. To assess the impact of this, the number of appearances of pictures of the red bus in our sample was counted; a high recall OCR
step was followed by a manual classification to find these images. 913 images of the bus were found. Furthermore, 21,240 appearances of the leave claim in some form of image were found, using a fully automated OCR method with an F1 of 0.87, substantially increasing the textual count for that claim. Moore and Ramsay~\shortcite{moore2017uk}
state that the remain claim was discussed in 365 newspaper articles,
whereas the leave claim was discussed in only 147. The greater media
interest in the Osborne claim is unsurprising given his position of
authority, but this didn't translate into interest on Twitter.

Note that not all Twitter discussion of the misleading headlines is
uncritical propagation. The tweets often talk about the credibility of
the headline. The 21,240 leave claim images were tweeted by 16,490 unique users. Of those, a higher number were remainers (5,369 vs. 4,950, with the remainder unclassified), suggesting a high proportion of Twitter interest in the claim was at least somewhat critical. Note also that although pictorial versions of the claim were tweeted by more remainers, the leavers that did tweet it tweeted it more; in terms of actual tweets containing pictures making the claim (buses as well as other imagery containing the claim) leavers accounted for 7531, compared with 6585 remainers, with the remainder unclassified, suggesting a greater enthusiasm for sharing the imagery among leavers, as one might expect. Recall that as we found above, our sample contains more remainers, but the leavers were more vocal. These findings recall Venturini~\shortcite{venturini2019}, who notes that the spreading of information is largely independent of whether the spreader actually believes it, and that this viral tendency and the resulting deluge of valueless information may be the more significant aspect of the problem. 
A similar result is found when considering another prominent pictorial campaign; the UK Independence Party's poster showing a large queue of people alongside the slogan ``Breaking Point'' and the suggestion that ``we must take back control of our borders''. The poster has been criticised for implying that the people in the poster are entering the UK as immigrants, whereas in fact the picture was taken in Slovenia~\footnote{\small{\url{https://www.theguardian.com/politics/2016/jun/16/nigel-farage-defends-ukip-breaking-point-poster-queue-of-migrants}}}. This claim was found in 3,388 tweets in pictorial form, of which leavers account for 948 and remainers, 1,007, the greater number, and the rest unclassified. In terms of unique users, 843 leavers posted the claim in image form and 890 remainers did so (1,331 unclassified). It is evident from the above that in this case, remainers repeated the leave claim more than leavers.

\section{Discussion}

We have presented evidence addressing the presence of partisan media, propaganda and post-truth politics in the run-up to the UK EU membership referendum on Twitter and in the media, as well as more broadly. With regards to partisanship in Brexit, we saw that websites linked in topically related tweets were most often neutral or bipartisan in their appeal. However, sources with \textbf{partisan} appeal also captured a sizeable portion of the debate, and of those, the leave-partisan materials were much more heavily propagated. Mainstream media with a stated remain stance produced materials appealing to both sides of the debate. Some mainstream media with a stated leave stance produced materials predominantly appealing to leavers.

A high degree of imbalance between leavers and remainers in those linking to a medium's website was found to suggest partisanship or even propaganda; materials with a strong appeal to leavers rather than remainers were plentiful and diverse, and included mainstream media and alternative media including US and other foreign sources. Materials with a strong appeal to remainers were fewer and less influential, and mainly comprised explicit campaign sites. Number of upheld press complaints correlates more strongly with a site's partisan appeal than the bias of the source as determined by the difference between its pro- and anti-Europe front pages (though both correlations are highly significant), suggesting that partisan appeal is capturing something other than the extent to which a source provides a voice for a particular opinion, and that misinformation may be a part of it. Evidence of Russian state involvement was modest. Automated accounts were in evidence.

The main evidence presented regarding \textbf{propaganda} was taken from a dataset identified by Twitter as originating in the Russian Internet Research Agency, an organization known to seek global influence through the dissemination of propaganda materials. Observation of this data suggests a learning process on their part regarding how impact can effectively be achieved. Tapping into deeply felt issues such as Black equality and patriotism has allowed a few skilled agents to build a large following, accounting for by far the greater part of the IRA's reach. The appetite of the audience for a particular message might therefore be seen as the ``Trojan Horse'', via which the desired message may then be insinuated. Indeed some difficulty may arise in distinguishing the vehicle message from the propagandistic message that motivates the efforts. A good vehicle may bide its time, or indeed be an end in itself (for example leading to financial benefit through advertising revenue).

Low effort approaches, such as possibly automated retweeting and large scale tweeting of pleasing but vague content, didn't appear to result in a high reach. One observed case of a fabricated scare fell entirely flat. Whilst success to very great extent is in the hands of a handful of highly skilled political trolls, scaling up reach beyond a few established popular accounts was achieved in conjunction with the events at Charlottesville in 2017. The material appears to be skilled and probably not automated. Future work exploiting this corpus should involve a deeper review of the Russian language IRA tweets. This would provide a greater understanding of the early history of an internet propaganda operation. Linked materials also provide more detailed material. The website ``ReportSecret.com'' has been highlighted above, along with other partisan press and alternative media in reference to the Brexit case. Furthermore the Russian accounts linked thousands of times to pages on the website LiveJournal, where extensive material more in the nature of personal opinion achieved a high reach; most-linked pages discuss the shooting down of Malaysian Airlines Flight 17, and are pro-Russian, anti-Ukraine. The material has provided an opportunity to benefit from the IRA's learning process in understanding how messages spread or fail to spread. However, the observations made here are preliminary only, and must form part of a more rigorous and complete picture formed of all available data, not just part, and backed up by controlled studies.

Claims made by leave and remain campaigns were reviewed in the context of \textbf{post-truth politics}. Echoing findings above, uptake of misleading leave claims was found to be high, dwarfing, for example, any evidence of Russian influence on Brexit. The greater hazard for public information may be the increasing tendency for public figures to take liberties with the truth.

A background issue through the findings is the issue of polarization. In the section on partisan media we found that the pro-remain Guardian newspaper attracted critical comment, which the Express did not do to the same extent, instead attracting upheld press complaints. This raises questions about the factors that encourage, or discourage, bipartisan discussion. Highly partisan materials were found to be evident in great quantities in the form of linked materials in the Brexit tweet sample. Whilst
these materials are of concern in that they are prolific and more often misleading, and are attracting significant attention, information consumers show a preference for linking more moderate materials, supporting previous research suggesting that there is a polarizing pull from those putting out their message on the internet. In the IRA materials we found that political trolls attracted the greatest following and achieved the greatest impact pushing at a small number of what might be seen as ``open doors''; topics where feelings are already running high. These existing cracks in society may offer opportunities for those that wish to create further division.

The release of the IRA dataset by Twitter is an important step forward in platforms working together with scientists to enable a better understanding of the new social dynamic they have created. Controversial posts and accounts are suspended at a very high rate, creating an issue for open and repeatable science on social media data. However the dataset was limited in that follower/followee networks weren't included. Gaining a full picture requires access to all related data, not only tweets from a particular set of accounts. Similarly the impact of retweets cannot be understood without information about the retweet rate of retweets. Fully understanding impact requires information about how often a tweet appeared on someone's screen. Moving forward requires a careful debate about privacy. Failing to have that debate may result in information being richly available to those with commercial objectives, namely the platforms themselves, but denied to a society reeling from the effects.

As already discussed above, disinformation and biased content reporting are not just the preserve of fake news and state-driven propaganda sites and social accounts. A significant amount also comes from media and factually incorrect statements by prominent politicians. The impact of widely known and influential claims made by politicians from both sides of the referendum campaign was already discussed above. Therefore, effectively combating deliberate online falsehoods must address such cases. Furthermore transparency in political advertising on social platforms and a review process for political advertising are likely to help with reducing the impact of all other kinds of disinformation already discussed above (i.e. fake news sites, Russian propaganda, etc).

\section*{Acknowledgments}


This work was partially supported by the European Union under grant agreements No. 687847 ``Comrades'' and No. 654024 ``SoBigData'', the UK Engineering and Physical Sciences Research Council grant EP/I004327/1 and the British Academy under call ``The Humanities and Social Sciences Tackling the UK’s International Challenges''.

\bibliographystyle{aaai}
\bibliography{influencers-brexit}

\vspace{2in}
\begin{table}
\begin{center}
\resizebox{.9\columnwidth}{!}{%
\begin{tabular}{|ll|ll|ll|}
    \multicolumn{6}{c}{\Huge{\textbf{Appendix}}}\\
    \multicolumn{6}{l}{}\\
    \multicolumn{6}{l}{}\\
  \hline
\textbf{Remain PAS\textgreater3:1} & \textbf{Total} & \textbf{Neutral} & \textbf{Total} & \textbf{Leave PAS\textgreater3:1} & \textbf{Total}\\
  \hline
gov.uk & 63119 & twitter.com & 4018371 & Youtube & 226382\\
theconversation.com & 8495 & The Guardian & 253474 & The Telegraph & 148565\\
internacional.elpais.com & 6915 & BBC & 242131 & Daily Mail & 86888\\
blogs.lse.ac.uk & 6532 & Facebook & 109552 & Bloomberg & 53071\\
jkrowling.com & 5975 & The Independent & 104572 & news.sky.com & 32016\\
economist.com & 5220 & amp.twimg.com & 80727 & The Sun & 30255\\
eureferendum.gov.uk & 4095 & Reuters & 71776 & snpy.tv & 28281\\
timeshighereducation.com & 3738 & wp.me & 58287 & Russia Today & 23064\\
politics.co.uk & 3344 & Financial Times & 44497 & cnn.it & 22617\\
politicalscrapbook.net & 3266 & mirror.co.uk & 43467 & on.wsj.com & 20332\\
secure.avaaz.org & 3159 & buff.ly & 40646 & itv.com & 17200\\
leftfootforward.org & 3014 & paper.li & 39458 & on.mktw.net & 16838\\
touchstoneblog.org.uk & 2655 & New York Times & 38441 & blogs.spectator.co.uk & 13298\\
zeit.de & 2476 & Huffington Post & 33697 & cnb.cx & 12946\\
snp.org & 2455 & econ.st & 29956 & forbes.com & 11967\\
tagesschau.de & 2396 & The Times & 25519 & yhoo.it & 7955\\
cer.org.uk & 2216 & cards.twitter.com & 21589 & Sputnik & 7032\\
greenpeace.org.uk & 2078 & standard.co.uk & 15335 & reportuk.org & 6712\\
lavanguardia.com & 2049 & instagram.com & 14671 & IBT & 6577\\
birminghammail.co.uk & 1856 & El Economista & 13665 & marketwatch.com & 6090\\
\hline
\end{tabular}
}
\end{center}
\captionof{table}{PAS\textgreater3:1 Sites and Sites with Neutral Appeal}
\label{tab:neutralishsites}

\vspace{20pt}

\begin{center}
\resizebox{.75\columnwidth}{!}{%
\begin{tabular}{|ll|ll|}
  \hline
\textbf{Remain PAS\textgreater30:1} & \textbf{Total} & \textbf{Leave PAS\textgreater30:1} & \textbf{Total}\\
\hline
ukstronger.in & 39221 & express.co.uk & 168846\\
prt.news & 20452 & breitbart.com & 55493\\
virg.in & 11708 & zerohedge.com & 20531\\
strongerin.co.uk & 10672 & Heat Street & 14889\\
infacts.org & 8165 & voteleavetakecontrol.org & 14235\\
ebx.sh & 4670 & order-order.com & 12804\\
voteremain.win & 2567 & infowars.com & 7306\\
unite4europe.org & 1554 & to-go.xyz & 6107\\
owl.li & 1462 & dld.bz & 5561\\
energydesk.greenpeace.org & 1169 & guyfawk.es & 5072\\
scotlandineurope.eu & 1166 & specc.ie & 4709\\
weareeurope.org.uk & 1151 & telegraaf.nl & 4659\\
realnewsuk.com & 1070 & dailysquib.co.uk & 4396\\
euromove.org.uk & 968 & davidicke.com & 4184\\
bmj.com & 900 & twibble.io & 4138\\
neweuropeans.net & 788 & brexitthemovie.com & 3997\\
greens.scot & 741 & eureferendum.com & 3673\\
richardcorbett.org.uk & 712 & au.news.yahoo.com & 3447\\
uktostay.eu & 696 & indiegogo.com & 3369\\
chokkablog.blogspot.co.uk & 691 & live.pollstation.com & 3269\\
\hline
\end{tabular}
}
\end{center}
\captionof{table}{PAS\textgreater30:1 Sites}
\label{tab:partisansites}

\vspace{20pt}

\begin{center}
\resizebox{.75\columnwidth}{!}{%
\begin{tabular}{|ll|ll|}
  \hline
\textbf{Left Trolls} & \textbf{Total} & \textbf{Right Trolls} & \textbf{Total}\\
\hline
twitter.com & 11132 & twitter.com & 2042\\
blackmattersus.com & 1788 & facebook.com & 593\\
blacktolive.org & 1572 & breitbart.com & 473\\
goo.gl & 844 & youtube.com & 419\\
instagram.com & 614 & washingtonpost.com & 275\\
youtube.com & 482 & foxnews.com & 165\\
tribpub.com & 279 & thegatewaypundit.com & 155\\
bb4sp.com & 261 & jihadwatch.org & 117\\
medium.com & 243 & dailymail.co.uk & 114\\
independent.co.uk & 237 & cnn.com & 112\\
huffingtonpost.co.uk & 230 & dailycaller.com & 92\\
theroot.com & 182 & medium.com & 91\\
vine.co & 176 & nytimes.com & 88\\
facebook.com & 143 & jennabrams.com & 63\\
rawstory.com & 140 & nypost.com & 63\\
thefreethoughtproject.com & 118 & infowars.com & 61\\
washingtonpost.com & 115 & vine.co & 60\\
tumblr.com & 98 & express.co.uk & 57\\
atlantablackstar.com & 94 & tribpub.com & 54\\
nytimes.com & 89 & thehill.com & 48\\
\hline
\end{tabular}
}
\end{center}
\captionof{table}{IRA Political Troll Most Linked Sites}
\label{tab:irasites}

\end{table}

\end{document}